# Bridging the Gap between Crosslinking Chemistry and Directed Assembly of Metasurfaces Using Electrohydrodynamic Flow

William J. Thrift,[1] Cuong Q. Nguyen,[1] Mahsa Darvishzadeh-Varcheie,[2] Nicholas Sharac,[3] Robert Sanderson,[4] Filippo Capolino,[2] Regina Ragan[1*]

[1] Department of Chemical Engineering and Materials Science, University of California, Irvine, Irvine, CA 92697-2575, USA

[2] Department of Electrical Engineering and Computer Science, University of California, Irvine, Irvine, CA 92697-2625, USA

[3] Department of Chemistry, University of California, Irvine, Irvine, CA 92697-2025, USA

[4] Department of Physics and Astronomy, University of California, Irvine, Irvine, CA 92697-4575, USA

*Corresponding author email:  rragan@uci.edu

## ABSTRACT

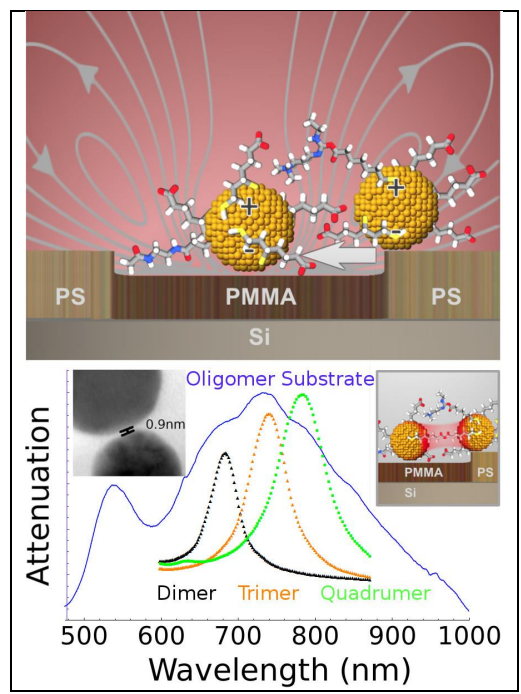

Advances in understanding chemical and physical driving forces in self-assembly allow the fabrication of unique nanoarchitectures with subwavelength building blocks as the basis for plasmonic and metamaterial devices. Chemical crosslinking of colloidal nanospheres has produced among the smallest gap spacings, necessary to obtain regions of greatly enhanced electric field, hotspots, which are critical to tailor light-matter interactions. However, obtaining uniform electromagnetic response of dense nanoantennas over large area for use in devices remains challenging. In this work, electrohydrodynamic (EHD) flow and chemical crosslinking is combined to form dense, yet discrete, Au nanosphere clusters (oligomers) on a working electrode. EHD provides a long range driving force to bring nanospheres together and anhydride crosslinking yields 0.9 nm gap spacings. Using selective chemistry, nanospheres are simultaneously crosslinked onto a block copolymer template, producing oligomers with a narrower wavelength band width and higher hotspot intensity than monolayer structures produced without a template. We investigate nanoantenna response via full wave simulations, ultraviolet-visible spectroscopy, and surface enhanced Raman scattering (SERS). Nanoantennas exhibit uniform hotspot intensity and gap spacing. Simulations show field enhancements of 600, correlating well with measured average SERS enhancement of $1.4 \times 10^9$. Nanoantenna substrates produce a SERS signal with a relative standard deviation of 10% measured over a 1 mm$^2$ area, crucial for nano-optical devices such as optical sensors, among other applications. Understanding long range (EHD flow) and short range (chemical crosslinking) driving forces provides the control for assembling colloidal nanoparticles in architectures for large area plasmonic and metasurface device fabrication.

# INTRODUCTION

Metal architectures using colloidal nanoparticles as meta-molecule building blocks have shown great promise as a scalable self-assembly method to control light matter interactions in large area devices. Such methods have produced ultrathin quarter wave plates,[1] perfect absorbers,[2] and ultrafast spontaneous emission sources.[3] The above applications can be realized with large area patterns or random deposition of nanoparticles on surfaces. By forming discrete sub-wavelength clusters, devices that rely on narrow-band resonances—i.e., Fano resonances—based on "dark" (i.e., low scattering) electric and magnetic resonances can be realized.[4–7] Natural magnetism fades away at infrared and optical frequencies, and conventional split-ring resonators, that in principle could provide narrow band resonances, are prohibitively difficult to scale down to optical wavelengths,[8] whereas coupled nanospheres can be scaled.[9,10] Surface enhanced Raman scattering sensors also exhibit better physical performance with discrete clusters rather than hexagonally close packed layers.[11,12] As these architectures are composed of sub-wavelength building blocks, traditional optical lithography methods cannot be utilized for large area device fabrication. Thus driving forces in colloidal assembly methods must be further understood to form discrete and dense assemblies of nanoparticles to fabricate metasurfaces. An important parameter to tune the optical response is the gap spacing between nanoparticles. When electromagnetic radiation impinges on nanoantenna with small gaps, i.e., hotspots, the electric field is greatly enhanced.[13] In metal nanoarchitectures, this enhancement is induced by localized surface plasmon resonance and increases with decreasing gap spacing of hotspots. Taking advantage of this, many nanoscale optical devices now rival the performance of bulk devices. While the smallest gap spacings, produced by electron beam lithography,[14] have reached – and surpassed – the quantum mechanical tunneling regime, approximately 0.5 nm, achieving gap spacings near this limit over large areas remains an on-going challenge for device fabrication. Thus devising new fabrication methods to arrange nanoparticles from colloids into discrete clusters will enable a new generation of compact optical devices.

Self-assembly using salt-induced aggregation produces small gap spacings,[15–18] yet such aggregation is difficult to control and results in large variability in both oligomer morphology and gap spacing. To solve some of these challenges, crosslinking chemistry has been introduced to initiate more control in nanoparticle linking and thereby gap spacing. Systems including DNA origami,[19–21] Cucurbit[n]uril,[22,23] organic molecules,[24–26] protein linkers,[27,28] and polymer encapsulation[29–31] have been shown to achieve small – even sub-nanometer – gap spacings. Chemically crosslinked nanoparticle oligomers have found applications in surface enhanced Raman spectroscopy (SERS),[32,33] surface enhanced fluorescence,[34] and surface enhanced circular dichroism.[35,36] While many of these systems are successful at producing oligomers in solution, depositing oligomers onto a substrate in a reproducible manner to generate uniform optical response over both large area and from sample to sample remains an ongoing challenge. Primary difficulties lie with achieving a high yield of oligomers that are small (field enhancement of large oligomers is reduced by loss), dense (yet discrete to avoid inter-oligomer coupling), have uniform gap spacing, and optically uniform (in terms of both size of probe, μm scale, and point to point, over cm scale distances).

To address these issues, directed self-assembly of nanospheres on templates fabricated using top-down fabrication has been developed.[37–40] While these methods have also found broad applicability in SERS,[41] second harmonic generation,[42] photon upconversion,[43] and projection lithography,[44] top-down fabrication of templates typically yields sparse architectures. In this work, we achieve large scale uniformity in hot spot intensity in densely packed Au nanosphere oligomers with sub-nanometer and uniform gap spacings by supplementing self-assembly on a chemically functionalized diblock copolymer template with both chemical crosslinking and electrohydrodynamic (EHD) flow.[45–48] EHD is a longer range driving force that facilitates chemical crosslinking. Induced by an electric field in colloidal solution, EHD flow has been shown to promote lateral motion and close-packing of particles in colloid at the electrode surfaces, assembling transient close-packed structures. EHD flow has been primarily studied in the context of micron scale particles to form monolayers. Here, we freeze in transient structures with local chemical crosslinking reactions and demonstrate a high yield of small oligomers and close packed oligomers composed of ten nanospheres or less with gap spacing of 0.9 nm. These oligomers exhibit high field enhancements with plasmon resonances in the range of 615 nm - 875 nm full width half maximum. Specifically, nanospheres are assembled using EDC, a carbodiimide crosslinker often used in bioconjugation,[49] polymer science,[50–52] and nanoparticle crosslinking[53,54] on a copolymer template composed of poly(styrene-b-methyl methacrylate) (PS-b-PMMA). The anhydride crosslinking pathway, one of the pathways of the carbodiimide chemistry, forms an anhydride bridge between the carboxylic acid ligands on Au nanospheres. The anhydride chemical bridges are observed through transmission electron microscopy to yield gap spacings of approximately 0.9 nm that are not observed in the absence of EHD flow. Furthermore, another pathway of the carbodiimide chemistry binds Au nanospheres to the PMMA domains on the copolymer template, yielding small and dense oligomers over the measured scale of 1 cm$^2$. Full wave simulations of fine structure in absorption measurements indicate ultraviolet-visible spectroscopy response is dominated by oligomers with gap spacings of 0.9 nm, indicating large area uniformity in gap spacing. Full wave simulations also show an electric field enhancement on the order of 600 in the hotspot region due to the narrow gap spacing in oligomers. The field enhancement correlates well with measured SERS enhancement factors, using benzenethiol as the analyte, are on average $1.4\times10^9$. The SERS signal exhibits a relative standard deviation of 10% across a 1 mm x 1 mm area with over 10000 different measurements, with measurements separated by 10 μm. These results demonstrate how assembly using EHD flow improves optical device performance as the contemporary state of the art SERS substrate have achieved variations of a factor of two (with enhancement of $10^7$)[55]. Overall, by understanding the influence of short range and long range driving forces in assembly from colloid using carbodiimide crosslinking chemistry combined with EHD flow, this work represents a step forward for directed assembly of nanoarchitectures to create metasurfaces by demonstrating the influence of both the template and physical driving forces in assembly to control oligomer morphology and uniformity of gap spacing.

## RESULTS AND DISCUSSION

### Nanoantenna Oligomer Deposition

Chemical assembly on self-organized templates allows for the formation of discrete oligomers over large areas that serve as nanoantennas. Block copolymers are especially desirable as templates for this method;[56–58] poly(styrene-b-methyl methacrylate) (PS-b-PMMA) has been used extensively to produce regular, nanometer-scale hotspots that enhance light-matter interactions.[11,12,54,56] Using a bioconjugation-inspired assembly process[59], the zero-length carbodiimide crosslinker EDC enables the assembly of Au nanospheres into discrete oligomers onto this self-organized chemical template. The copolymer minimizes interfacial energy between the two polymer blocks by self-organizing into PMMA domains with widths of approximately 40 nm separated by PS. These PMMA domains are selectively functionalized with ethylenediamine to form an amine surface group as the PS regions are chemically inert in the ethylenediamine solution. Carboxylic acid terminated Au nanospheres are linked to the amine functionalized PMMA domains, left inset of Figure 1a, via peptide bond formation through consecutive reactions with EDC, right inset of Figure 1a. An oligomer is formed when a consecutive Au nanosphere surface linkage occurs near an existing nanosphere. Oligomer formation is promoted through the use of one of two driving forces: Brownian motion (zero bias-voltage control) and electrophoretic deposition (EPD), as illustrated in Figure 1a and 1b, respectively. In the case of zero bias-voltage deposition, a copolymer-coated Si substrate is placed in a colloidal Au nanosphere solution and Brownian motion drives the diffusion of Au nanospheres to collide with the substrate surface. In the second deposition method, EPD, an applied electric field drives the Au nanospheres toward the copolymer surface, a process termed electrophoretic sedimentation. Additionally, the applied potential generates EHD flow.[45] A local inhomogeneous electric field is generated due to the polarization of the Au nanospheres on the copolymer-coated Si electrode surface. This field in turn drives ion motion, and thus fluid flow.[60] The net result is a lateral attractive force between nanospheres at the colloid electrode interface that causes the nanospheres to aggregate. Figure 1b includes a schematic of this flow field in the inset. During EPD, the electric field leads to electrolysis that reduces the pH at electrode surfaces, and thus minimizes the electrostatic repulsion between carboxylate groups on the Au nanospheres. EPD deposition is performed in two 10-minute deposition cycles at 60°C. A copolymer-coated Si substrate and Pt mesh are used as the working electrode and counter electrode, respectively. For each deposition cycle, fresh nanosphere solution is used to prevent aggregation. While oligomers are produced with both methods, EPD has been observed to increase the formation of oligomers and reduces the gap spacings.[11] Yet the physical mechanisms driving oligomer formation with narrow gap spacing have not been previously reported.

Figure 1. Schematic of experimental setup where colloidal solutions of Au nanospheres with lipoic acid ligands are chemically assembled on self-organized PS-b-PMMA diblock copolymer templates, see inset of (a). Two different

driving forces to bring nanospheres to the surface were investigated: (a) zero bias-voltage control deposition that relies on Brownian motion and (b) electrophoretic deposition that leads to electrohydrodynamic (EHD) flow, inset, on the template surface.

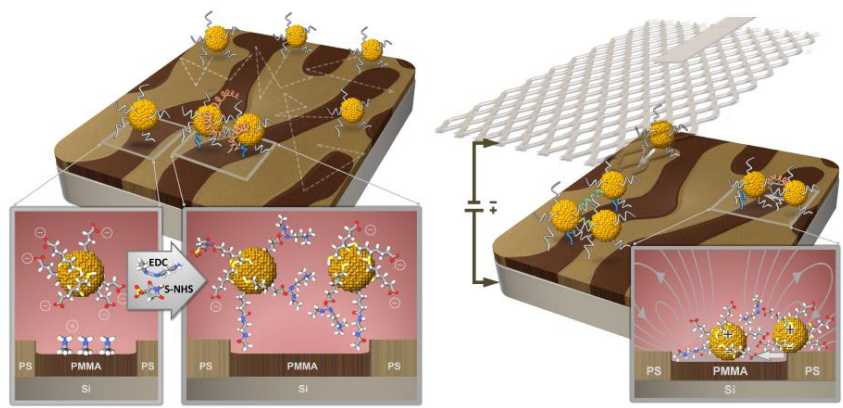

The gap spacings observed in this work are remarkable; among the smallest ever achieved over large area, and approaching the quantum tunneling limit.[14] Gap spacings are measured with a JEOL 2010F STEM using high angle annular dark-field imaging in scanning transmission electron microscopy (STEM) mode. Figure 2a and Figure 2b show transmission electron microscopy (TEM) images of 20 nm Au nanosphere dimers deposited using the zero bias-voltage control deposition and EPD, respectively. TEM imaging of single oligomers requires lower nanosphere coverage, so deposition time is limited to 5 minutes. Other deposition parameters remain as reported above, albeit on copper TEM grids instead of Si substrates. The resolution of the TEM images is reduced by the 40 nm thick layer of PS-b-PMMA that the beam must travel through to reach the detector. Zero bias-voltage control deposition results in gap spacing as low as approximately 2 nm, which falls into the 2-7 nm range previously reported.[11] The observed 2 nm gap spacing in the TEM image of Figure 2a corresponds to the length of two lipoic acid-stabilizing ligands. EPD, on the other hand, produces gap spacing of approximately 0.9 nm, shown in Figure 2b. Surprisingly, the measured EPD gap spacing of 0.9 nm is shorter than the length of two lipoic acid ligands. While narrow, these gap spacings are still large enough to avoid the depolarization of their dipolar resonances due to quantum tunneling, which has been shown to reduce coupling efficiency when gap spacing is smaller than 0.5 nm.[14]

Figure 2. Transmission electron microscopy images of Au nanospheres deposited on Cu TEM grids coated with PS-b-PMMA using (a) zero bias-voltage control and (b) electrophoretic deposition.

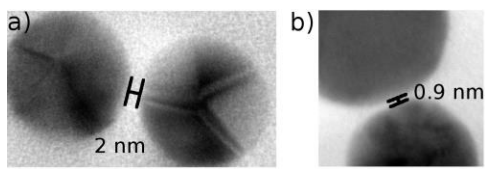

## Influence of EHD Flow on Oligomer Morphology

The importance of EHD flow for nanometer scale particles and other physical parameters in the deposition process were investigated to probe their influence on gap spacing, areal density, and geometry in oligomer formation. Use of larger diameter nanospheres unexpectedly led to substantial differences in oligomerization (defined as the tendency to form oligomers with an increased number of Au nanospheres), density, and morphology. Oligomers are observed using scanning electron microscopy (SEM) images acquired with a Magellan XHR SEM (FEI). SEM images are analyzed using Wolfram Mathematica[TM] (described in the supplemental information) to examine oligomerization, density and morphology of the nanoantenna (Au nanosphere) samples. Figure 3a plots the average oligomer distribution statistics of nanoantenna samples as a function of deposition parameters and nanospheres diameter. Representative SEM images of the sample surfaces are plotted in Figure 3c-g, and Figure S3.

The average and standard deviation of percent coverage of oligomers on samples with 20 nm and 40 nm Au nanospheres are determined from 25 adjacent regions with dimensions of 4.0 μm × 2.7 μm and 8.0 μm × 5.5 μm for samples, respectively. Percent coverage is useful for direct comparison of oligomers with different monomer diameter as studied here. Note that the area under the curve for the 40 nm EPD sample is 15 percent coverage which corresponds to 54 percent of the surface area of PMMA domains.[11] The relative calculated oligomer densities per micrometer squared are listed in Table 1.

Figure 3. (a) Oligomer distribution statistics, reported as percent coverage, observed on nanoantenna samples as a function of nanospheres size and deposition parameters. (b) Schematic illustrating carbodiimide crosslinking chemical pathways (c)-(g) SEM images of nanoantenna samples with different deposition parameters

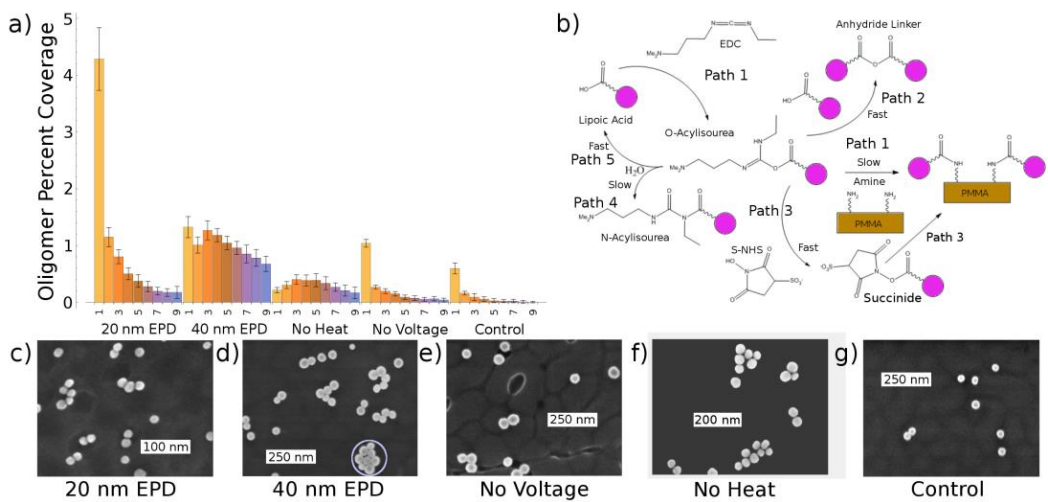

Table 1. Summary of the oligomer number density ratio (expressed as a percentage) between dimers and monomers (single nanospheres), trimers and monomers, and quadrumers and monomers. The average number of nearest neighbors for a given nanoparticle is also listed. Ratios are determined from analysis of 25 different SEM images having dimensions of 4.0 μm × 2.7 μm (20 nm) and 25  8.0 μm × 5.5 μm (40 nm).

|  | EPD 20 nm | EPD 40 nm | No Heat | No Voltage | Control |
| --- | --- | --- | --- | --- | --- |
| Total Coverage | 10.4% (1.7% σ) | 14.6% (1.5% σ) | 3.7% (0.2% σ) | 2.5% (0.9% σ) | 1.1% (0.3% σ) |
| Dimer: Monomer Ratio | 13.2% (1.6% σ) | 36.7% (4.9% σ) | 62.7% (19.7% σ) | 11.7% (1.7% σ) | 12.7% (3.2% σ) |
| Trimer: Monomer Ratio | 5.9% (0.9% σ) | 29.2% (3.7% σ) | 55.0% (18.0% σ) | 5.7% (1.2% σ) | 4.6% (2.6% σ) |
| Quadrumer: Monomer Ratio | 2.8% (0.7% σ) | 20.3% (2.8% σ) | 40.2% (14.3% σ) | 3.3% (0.9% σ) | 2.0% (1.5% σ) |
| Average Number of Nearest Neighbors | 1.21 | 2.06 | 2.23 | 1.19 | 0.66 |

Variation in nanosphere radii elucidates physical and chemical mechanisms that affect oligomerization. First, we observe a larger area under the oligomer distribution curve for the 40 nm samples compared to the 20 nm samples – both fabricated using EPD – indicating a greater tendency for 40nm Au spheres to bind to the PMMA surface. One factor that will lead to higher chemical attachment efficiency is the lower curvature of 40 nm versus 20 nm spheres, as lower curvature will relatively increase the number of O-acylisourea groups in contact with the amine-terminated PMMA during a collision. Yet, it is not just surface coverage that increases with increasing nanospheres radii. As tabulated in Table 1, the 40 nm EPD sample exhibits nearly double the number of nearest neighbors to a given particle, 2.06, than the 20 nm EPD sample, 1.21. This figure of merit provides information on the strength of the driving force to form oligomers on a sample, where 0 would be a monomer, and 6 would be a perfect hexagonally close packed lattice. This result is surprising as one might assume that the aggregation of Au nanospheres into oligomers is driven by Van der Waal's forces overcoming electric double layer forces, as described by Derjaguin Landau Verwey Overbeek (DLVO) theory.[61] DLVO theory predicts a higher degree of aggregation with smaller particles due to the electric double layer forces scaling with particle radii.[62] Additionally, a low pH near the working electrode is also expected to reduce the electric double layer, further promoting the reversible aggregation of lipoic acid-functionalized particles near the substrate surface. However, this is not consistent with the data, as the 40 nm nanosphere samples exhibit a higher degree of oligomerization as observed by the oligomer-to-monomer ratios and average nearest neighbor count listed in Table 1. Consequently, one can conclude that the oligomerization of the nanospheres is driven by a different mechanism than thinning of the double layer. Furthermore, 3-dimensional aggregates and large fractal structured aggregates, characteristic of aggregation in solution, are not commonly observed in SEM images, see Figure S2a. Thus EHD flow appears to play an important role in the formation of oligomers on the chemical template during EPD; this is beyond the expected role of providing a driving force to bring nanospheres toward the surface via electrophoretic sedimentation due to an applied potential on the sample surface. In addition, thinning of the double layer does not appear to be a strong driving force in oligomerization.

Table 2. Summary of experimental conditions used in the deposition of the various samples.

|  | 20 nm EPD | 40 nm EPD | No Heat | No Voltage | Control |
|---|---|---|---|---|---|
| Nanosphere Diameter | 20 nm | 40 nm | 40 nm | 40 nm | 40 nm |
| Applied Voltage | 1.2 V | 1.2 V | 1.2 V | 0 V | 0 V |
| Temperature | 60 $^{0}$C | 60 $^{0}$C | 20 $^{0}$C | 60 $^{0}$C | 20 $^{0}$C |

In order to further examine the importance of EHD flow in governing nanoantenna architecture, it is important to note that samples fabricated using EPD also involves an elevated processing temperature of 60°C. To decouple the role of EHD flow and that of temperature, we evaluate architectures arising from the following experimental parameters, tabulated in Table 2: zero bias-voltage at T = 60 °C, room temperature deposition at an external bias of V = 1.2V, and zero bias-voltage and room temperature deposition, labeled 'No Voltage,' 'No Heat,' and 'Control' in Figure 3a, respectively. The oligomer distribution statistics of the control nanoantenna sample, in Figure 3a, unsurprisingly exhibits few nanospheres on the surface. The lower temperature appears to reduce Brownian motion, leading to fewer collisions between nanospheres and functionalized PMMA regions and thereby lowers surface coverage. The measured statistics for the control sample also show virtually no oligomers over size 4. However, comparison of the surface coverage of the No Voltage and No Heat samples shows that an applied bias has an even greater impact on surface coverage than temperature. It is also worth noting that the oligomer:monomer ratios of the No Heat samples are the largest, albeit at the expense of uniformity as measured by the standard deviation. The greater degree of oligomerization is clearly due to the bias, either through EHD flow or electrophoretic sedimentation.

The data shows that electrophoretic sedimentation is apparently not the primary driving force for attachment for samples fabricated with an applied voltage. While the greater frequency of monomers on the 20 nm EPD sample compared to all other samples appears to be an effect of electrophoretic sedimentation, the smaller radii reduces stokes drag which increases sedimentation, when we keep radii constant and vary the applied bias-voltage we observe a different effect. Applying a bias of 1.2V at room temperature actually leads to over a 2.5-fold reduction in the number of monomers. This data indicates that for the 40 nm particles the EHD flow is more impactful on the oligomerization compared to electrosedimentation.

Another observation deserves mention: oligomers formed with a bias voltage are not necessarily aligned with the underlying PMMA domains. In the No Voltage and Control samples, we observe the nanospheres directly above PMMA domains, as expected when carbodiimide crosslinking between the sphere's lipoic acid ligands and the amine-functionalized PMMA is the main attachment mechanism. The situation is very different for the 40 nm EPD sample and especially the No Heat samples. Here, we observe oligomerized nanospheres directly over PS domains, apparently not attached to the PMMA, but to another nanosphere, an example of this is shown in Figure. 4d where the oligomer is circled in blue. This, and the observation that temperature has a profound effect on oligomerization, as shown by the comparison between the No Heat sample and the 40 nm EPD sample, demonstrates that chemical crosslinking also plays a significant role.

This leads us to reexamine the carbodiimide crosslinking reaction, shown in Figure 3b, as this provides the final needed insight into how the nanosphere radii and deposition process drives oligomerization. As illustrated and labeled Path 1 in Figure 3b, EDC is used to crosslink carboxylic acid end groups on nanospheres ligands and primary amines on the PMMA template. This mechanism forms oligomers when a nanosphere chemically binds next to an existing monomer on a PMMA domain, resulting in oligomers that are aligned with the PMMA domain. However, if carbodiimide crosslinking is the main driving force of assembly, one would expect a large number of monomers on the surface with a monotonically decreasing occurrence frequency of larger oligomers as electrostatic repulsion between nanospheres would favor monomer formation. While we observe a large fraction of monomers in the No Voltage and Control samples, the 40 nm EPD and No Heat samples exhibit a preference for oligomers. This indicates that standard carbodiimide crosslinking is not strongly driving oligomer formation. During the deposition process EDC and s-NHS are used; EDC reacts with carboxylic acid end groups, first to form O-acylisourea esters and then to form peptide bonds, and s-NHS is added to consume reactive O-acylisourea esters on the surface of the Au nanospheres (Path 3 in Figure 3b) preventing the formation of N-acylisourea, labeled Path 4, that leads to unreactive nanospheres. However, the relatively small concentration of EDC/S-NHS and the ratio of 1:2.5 used here also promotes the formation of anhydride groups that bridge carboxylate groups on the nanospheres[50] (Path 2). The measured EPD gap spacing of ~0.9 nm in TEM images is consistent with the length of an anhydride bridge, where the sulfur-sulfur distance of the anhydride bridge is predicted to be 0.86nm by PyMol$^{TM}$. While this reversible reaction can occur in solution, we observe no permanent 3-dimensional aggregates that would arise from spontaneous anhydride bridging in colloidal solution in Figure S2a, which displays an SEM image over 50 μm x 50 μm area. This indicates that the attractive force between nanospheres is present on the surface of the working electrode. Thus formation of planar oligomers on nanoantenna surfaces involves a 2-dimensional driving force.

The force induced from EHD flow and the resultant anhydride bridging between nanospheres has all of the ingredients necessary to explain the complete oligomerization mechanism. EHD flow is a 2-dimensional attractive driving force that occurs at the electrode surface[60] and drives nanospheres into close packed arrangements at the substrate surface to reduce free energy[63]. When nanosphere's carboxylic acid end groups are brought close together, the formation of anhydride bridges can occur and this will lock in transient structures that occur due to EHD flow. EDC has previously been observed to form anhydride bridges between carboxylic acid groups during polymer crosslinking.[52] Oligomers produced via anhdydride bridging (Path 2) will not be fully commensurate with the PMMA domains as oligomers forming from carbodiimide crosslinking (Path 1) on PMMA alone. With electrohydrodynamically driven bridging, oligomerization is promoted through the assembly of nanospheres around a monomer already attached to PMMA. This allows for the formation of close-packed oligomers, such as the circled oligomer in Figure 4d. An increase in average number of nearest neighbors in samples fabricated with an applied voltage as shown in Table 1. In addition, the EHD force increases with increasing particle size[48] consistent with the data showing a greater tendency for oligomerization is observed with increasing nanospheres radius. Yet even in the presence of EHD forces, oligomerization is still affected by carbodimiide crosslinking as monomers chemically bound to PMMA domains serve as the perturbations to the local electrostatic potential that is the origin of the EHD flow itself. These monomers then act as nucleation sites for the larger oligomers to form, attracting additional particles via EHD flow, and linking them with anhydride bridges. The ability for EHD flow to drive the formation of anhydride bridges between Au nanospheres in oligomers has profound implications on the uniformity of the gap spacings in the oligomers and their optical properties.

## Oligomer Optical Properties

The uniformity of the optical response of nanoantenna samples is investigated in a Shimadzu UV-1700 absorption spectrometer. A 9 mm x 0.5 mm area of a sample is illuminated, probing millions of oligomers to determine the aggregate optical response of a sample. Nanoantenna samples are fabricated on transparent substrates by spin coating the PS-b-PMMA copolymer template onto indium tin oxide (ITO) coated glass, as detailed in the methods section. Figure 4a,b shows the normalized attenuation spectra of a 40 nm EPD sample and a 40 nm No Voltage sample, respectively. Measurements are performed on samples immersed in water to simulate the environment in a biosensing experiment. The monomer resonance is observed in both spectra at 536 nm, in good agreement with Mie scattering theory. The oligomer resonances, observed from 625 nm to 875 nm, result from the different linear oligomer geometries and degree of oligomerization. The fine structure of the attenuation spectra is elucidated via comparison to the calculated oligomer absorption cross section. We determine the absorption cross section of representative oligomer geometries with full-wave simulations using the finite elements method (implemented via CST Microwave Studio), shown in Figure 4a and detailed in the methods section. Simulations of the oligomer resonances are performed using the experimentally observed gap spacings of 0.9 nm (also the calculated length of an anhydride bridge). These simulations use 2.47 for the permittivity of the surrounding BZT and PMMA layers, and depict the nanospheres as embedded into the PMMA by 30%, as previously reported.[12] The permittivity in the gap region is uncertain as the excitation source will probe a volume composed of the anhydride linker, aqueous solution, and copolymer environment. In order to account for this, a parameter sweep of the gap permittivity is performed using the dimer configuration, and the permittivity which best corresponds to the observed fine structure peak at 686 nm is determined to be 2.25. This parameter is reasonable based on an estimate of the average permittivity of the materials[12] and further verified by the correspondence between the trimer configuration's simulated plasmon resonance and the second fine structure peak observed in the 40 nm EPD sample. The overlap between the simulated absorption cross-section and measured peaks indicates the fine structure in the measurements is dominated by resonances of oligomers composed of two to four nanospheres having 0.9 nm gap spacing. These uniform 0.9 nm gap spacings across the oligomers samples result in extraordinary field enhancements, shown in Figure S3, which exceed 600.

Figure 4. Normalized attenuation spectra of a (a) 40 nm EPD sample (solid blue curve) and (b) absorption by a sample surface assembled without voltage (No Voltage sample, solid red curve). Calculated absorption cross section of oligomers with various geometries (dimer: black triangles, linear trimer: orange triangles, linear quadrumer: green squares, close packed quadrumer: dashed yellow curve, close packed hexamer: solid brown curve, close packed octamer: dot-dashed purple curve) is overlaid on data in (a). Close packed absorption cross-sections offset by 39000 for visual clarity. In (b) the calculated maximum value of absorption cross section of a linear dimer, trimer and quadrumer are shown with a black, red, and green lines, respectively. (c) The schematic of dimer structure used in full wave simulations; the layer of analyte and ligands has thickness, $t = 0.9$ nm, and permittivity of 2.47. The region in the gap has height, $l_x = 13.4$ nm, and gap spacing of 0.9 nm. Except in the case of the 2 nm gap dimer, whose resonance wavelength is shown as a dashed magenta line in (b) The permittivity of the gap region, $\varepsilon_x = 2.25$, was found in simulations via a parameter sweep to match attenuation spectra.

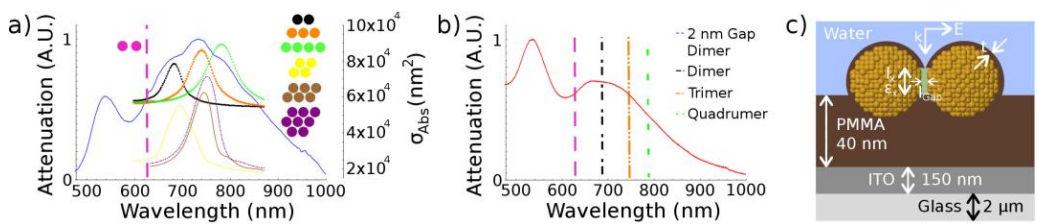

The impressive correspondence between simulation and observation absorption features for the 40 nm samples elucidates the effect of EHD flow on the self-assembly of oligomers. First, we observe good contrast between peaks of varying degrees of oligomerization. The narrow dimer resonance in the 40 nm EPD sample then demonstrates that most dimers are connected via anhydride bridges. The dashed magenta line in Figure 4a,b depicts the plasmon resonance of a dimer with a 2 nm gap spacing, this configuration has a calculated resonant frequency of 636 nm, where the 40 nm EPD sample shows insignificant attenuation at this wavelength. Indeed, we observe a broader and blueshifted dimer resonance corresponding to a 2 nm gap spacing in the No Voltage sample, where there is no expected EHD flow to promote the formation and retention of these anhydride bridges. This is in agreement with the separation between Au nanospheres observed in TEM and SEM images for samples fabricated in the absence of an applied voltage.

Given the observation that EHD flow drives the close packing of Au nanospheres, we expect to see a contribution of these close-packed oligomers on the optical attenuation of our samples. Dark field microscopy measurements[64] and FDTD simulations[65] have previously shown that the addition of a nanosphere on an oligomer that is not in the direction of the incident beam's polarization yields a plasmon mode that is only slightly perturbed from the original oligomer. Simulations of a close packed quadrumer, which has two particles lying along the polarization direction, show a resonances slightly red shifted compared to that of the dimer oligomer (yellow curve in Figure 4a). Similarly, the close packed hexamer (brown curve in Figure 4a) and octamer (purple curve in Figure 4a), which have three particles lying along the polarization direction show a resonance slightly red shifted compared to that of the trimer oligomer. The first perfectly close packed oligomer to have greater than four particles in a row contains 21 particles; 98.4% of oligomers observed in the 40 nm EPD sample contain fewer than 21 particles. Furthermore, the distribution statistics by number show that 93% of the surface is composed of oligomers of nine nanospheres or less. This is the genesis of the relatively narrow optical response of the sample; while there is a distribution of oligomer size, the uniform gap spacings and the close-packed geometries produce plasmon resonances that are all very similar. The similarity of the resonances of these different oligomers results in enormous and uniform field enhancements over a large area when excited by the same wavelength laser. This enormous enhancement is in the gap region and slowly fades away from it.

**Characterization of Electric Field Enhancement Via Surface Enhanced Raman Scattering**

Next, the electric field enhancements provided by nanoantenna oligomers is investigated via surface enhanced Raman scattering (SERS). SERS is measured by depositing a standard analyte (self-assembled monolayer) on a nanoantenna sample and observing the inelastic scattering of light – termed Raman scattering – with an energy shift that is associated with the vibrational modes present in the analyte. Here, the signal enhancement is proportional to the electric field enhancement approximatively taken to the fourth power, making SERS extremely sensitive to the performance of the antenna sample. Samples are first assembled as described in the methods section, treated with base (discussed below), and then treated with benzenethiol (BZT); the SERS spectra are obtained with the Renishaw InVia Raman spectrometer and measurements are detailed in the methods section. Figure 5a depicts the Raman scattering spectra obtained with BZT analyte on the nanoantenna samples and of neat BZT solution. As outlined in the supplemental information, calculations of the number of molecules in the sampling volume in the neat measurement of the BZT solution contains seven to eight orders of magnitude more molecules than SERS measurements of the nanoantenna samples. This large disparity is observed because a 3d volume is probed for the neat measurements; while only a 2d self-assembled monolayer of BZT on the nanoantenna regions is probed for our samples. Thus, qualitatively, the nanoantenna sample's comparable magnitude of Raman scattering intensity clearly conveys the large electric field enhancement due to our nanoantenna oligomers.

Figure 5. (a) Raman and SERS intensity of vibrational modes of BZT. The spectra displayed are as follows: neat solution (green curve, offset by 2000), No Voltage SERS spectra (blue curve offset by 8000), 20 nm EPD SERS spectra (purple curve offset by 16000 and normalized to account for the different CCD used), No Heat SERS spectra (orange curve, offset by 30000), and 40 nm EPD SERS spectra (red curve offset by 45000) all offsets are for visual clarity. (b) Calculated enhancement factor for the 998 cm$^{-1}$ and 1573 cm$^{-1}$ vibrational modes of BZT of the various samples.

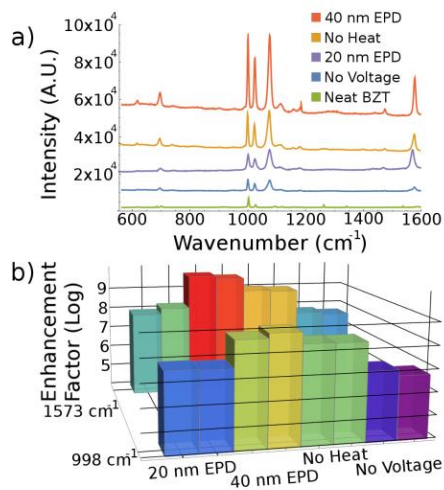

Figure 5b quantifies the performance of the nanoantenna oligomer sample through the sample's enhancement factor (EF). We define EF as the SERS intensity from the sample ($I_{SERS}$) divided by the neat Raman intensity ($I_{neat}$) in solution normalized by laser power and exposure time,[66,67]

$$EF = (I_{SERS}/N_{SERS})/(I_{neat}/N_{neat}) \qquad (1)$$

where $N_{SERS}$, $N_{neat}$, are the number of molecules participating in the SERS and neat measurements, respectively. Thus the SERS EF is the ratio of Raman scattered intensity per molecule of the antenna sample to that of the neat analyte. See supplemental information for the full derivation of EF. Figure 5b shows the calculated EF taken from

No Voltage, No Heat and EPD nanoantenna samples for both the 998 cm$^{-1}$ Ring out-of-plane deformation and 1573 cm$^{-1}$ C–H out-of-plane bending peaks. Raman scattering peaks from two different vibrational modes are used to calculate EF as the Raman scattering intensity may vary between vibrational modes for several reasons, including resonant enhancement due to charge-transfer transitions and other chemical enhancement pathways.[68]

The 40 nm EPD sample produces an EF as large at $3\times10^9$ using the 1536 cm$^{-1}$ Raman band, outperforming all other samples. First, the 40 nm EPD sample outperforms the 20 nm EPD sample two orders of magnitude. This is due to four factors: (1) the additional anhydride bridging in the 40 nm case due to increased force density generated by EHD flow;[48] (2) the intrinsically higher electric field enhancements that results from 40 nm nanospheres, observed in the simulation of a dimer in Figure S3; (3) the 40 nm oligomers, of all sizes, are closer to resonance with the 785 nm excitation wavelength compared to the larger oligomers in the 20 nm case which are off resonance at the 633 nm excitation; (4) a greater number of occupation sites for BZT in the hotspot between 40 nm nanospheres than in the 20 nm case, which scales with surface area. Next, the 40 nm EPD sample also outperforms the No Heat sample by nearly an order of magnitude. There is a greater fraction of oligomers composed of five or more nanospheres in the No Heat sample composed in a linear arrangement, such structures are off resonance at 785 nm laser excitation. Consider that increased temperature increases Brownian motion, which in turn promotes monomer formation via increased collisions of Au nanospheres with the copolymer template surface. These monomers lead to EHD flow as outlined above, and promote the further attachment of Au nanospheres to the existing Au nanospheres. This growth is limited by jamming of the surface by other existing oligomers, so increased monomer deposition will prevent large oligomers from growing. This process is analogous to crystal growth, where a high density of nucleation sites leads to the formation of smaller grains. Thus one would expect the No Heat sample to have fewer monomers serving as 'nucleation' sites and thus larger oligomers will form in comparison to the 40 nm EPD sample. The No Voltage sample achieves an EF of $5\times10^7$, this large enhancement factor corresponds to a >2 nm gap spacing in full wave simulations.[12] Overall, the SERS data reaffirms the cluster statistics presented in Figure 3, showing that the EHD flow has a profound impact on the formation and retention of anhydride bridges between nanospheres.

Figure 6. (a) Schematical representation of the 30% NH$_4$OH solution base treatment and BZT addition to oligomer substrates, depicting the increase of hot spot occupation by BZT after the anhydride bridge is cleaved. (b) SERS intensity of the 40 nm EPD sample before (blue curve) and after (orange curve) base treatment to cleave the anhydride bridge. The 40 nm EPD sample after base treatment curve is offset by 30000 for visual clarity.

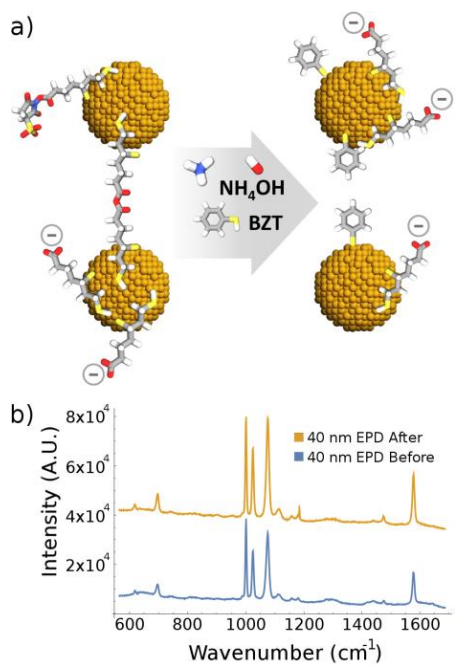

While anhydride bridging between Au nanospheres is useful to achieve narrow and uniform gaps between nanospheres, it may prevent an analyte from residing in a hot spot. We hydrolyze the anhydride groups,[69] illustrated in Figure 7a, by treating the nanoantenna oligomer samples with 30% NH$_4$OH solution for 90 minutes, with a 1 minute DI water rinse every 30 minutes. Figure 7b depicts SERS spectra of BZT from a 40 nm EPD sample, before and after the NH$_4$OH base treatment. The spectra taken are from identical locations which are located by bleaching a 8 μm × 5.5 μm copolymer area through electron bombardment with SEM. A 145% improvement in signal of the 998 cm$^{-1}$ band is observed for the 40 nm EPD sample. The improvement in signal is due to two mechanisms arising from cleaving the anhydride bridging: (1) the reduction of charge transfer plasmon modes that are non-radiative and do not contribute to the Raman signal and (2) freeing up occupation spots in the hotspot between nanospheres for BZT to occupy. Mechanism (1) is explained by the increase of the 998cm$^{-1}$ peak. However, mechanism (2) requires further analysis of individual peaks in the Raman spectra before and after hydrolysis of the anhydride bridge. After treatment, the 1076cm$^{-1}$ peak – in-plane ring deformation superimposed with C-S stretch - increases in strength relative to the 998cm$^{-1}$ peak - in-plane ring deformation. This suggests a change in the orientation of the BZT in the hotspot after base treatment, which alters the relative intensity of different Raman peaks. The 162% improvement of the 1076 cm$^{-1}$ mode – in comparison to 145% for 998cm$^{-1}$ mode – implies that the C-S stretch vibration is enhanced more than the ring breathing vibration after base treatment. This indicates that Au-S bonds displace the anhydride bonds previously located directly between the Au nanospheres where the electric field is the largest. Furthermore, one observes that the 634cm$^{-1}$ and 667cm$^{-1}$ peaks no longer appear in the spectra after base treatment. Vibrations of this energy are often C-H wagging modes; their disappearance suggests that they are C-H wagging modes of the anhydride bridge. These results further demonstrate that anhydride bridges form during oligomerization as described above; understanding this process leads to practical improvement of the sample's performance as an optical sensor.

Figure 7. SERS EF map across a 1 mm x 1 mm area where each measurement is separated by 10 μm, corresponding to 10000 measurements. The laser power is 76 μW with a 0.1s exposure time. Inset: histogram of measured SERS EF values with 57 bins depicting a mean of 1.4×10$^9$, FWHM of 3.5×10$^8$, and RSD of 10%.

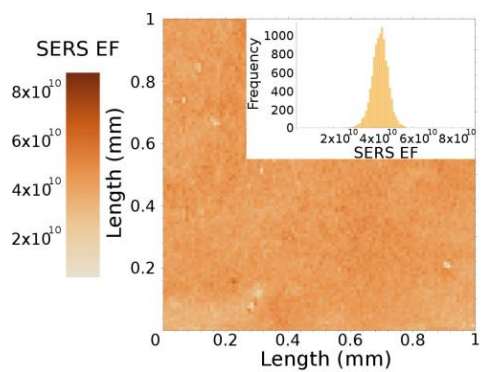

The large scale uniformity of the 40 nm EPD nanoantenna samples is demonstrated by acquiring a 1 mm × 1 mm SERS intensity map and calculating EF for each measurement, depicted in Figure 7. SERS measurements are separated by 10 μm in each direction. The raw SERS data is converted into an EF using equation 1 where the average surface coverage over the mapped region is obtained using SEM images. The average is used for all EF calculations due to practical difficulties with aligning the Raman map and the SEM map. The SERS map shows an average SERS EF value of 1.4×10$^9$. The data shows a 10% relative standard deviation (RSD) of the 1573 cm$^{-1}$ peak across the mapped area with a FWHM of 3.5x10$^8$, which is less than the proposed variability for an acceptable spot to spot variation in an effective SERS sensor.[70] These results demonstrate that the uniform oligomer deposition, demonstrated in the statistics shown in Figure 3 and the narrow optical response of the sample shown in Figure 4a, remarkably lead to highly uniform electric field enhancements across extremely large areas. Measurements taken with a 76 μW at 0.1s exposure on the 40nm EPD nanoantenna samples still have an excellent signal to noise ratio. This is extremely useful for diagnostic studies requiring fast analysis and low power, as is often the case in biosensing applications. It should also be noted that all SERS samples were reusable, and have a shelf life on the order of years. Re-immersing the nanoantenna samples in benzenethiol solution every few days yields the same SERS results.

## CONCLUSION

In this work, we have demonstrated the importance of combining short range and long range driving forces in self-assembly of Au nanospheres from colloid. Nanoantenna surfaces composed of oligomers are formed using local chemical reactions and EHD flow that drives nanospheres together for the formation and retention of anhydride bridges. This chemical crosslinking fabrication in the presence of an electric field produces large area hotspots with a uniform gap spacing of just 0.9 nm, a length scale ideal for field enhancement as it is just larger than the quantum mechanical limit for charge transfer plasmon modes leading to depolarization. The combination of the PS-b-PMMA chemical template, EDC cross linking chemistry, and EHD flow overcomes many of the challenges associated with colloidal deposition, namely low oligomer density and loss associated with large oligomers. Furthermore, we show the templated, EHD driven chemical crosslinking produces dense oligomers with a narrow optical response. Full wave simulations and ultraviolet-visible spectroscopy demonstrate that the vast majority of oligomers on nanoantenna surfaces are excited by a 785 nm excitation source. We further demonstrate SERS EF average over $1.4 \times 10^9$ over a 1 mm × 1 mm area, one of the largest ever reported for a large area self-assembled SERS sensor having a low relative standard deviation of 10% in signal intensity. Such low relative standard deviations are essential for metabolomics studies of small molecules, as a 10-fold increase in concentration often leads to significant increases in measured intensity of an analyte's Raman band.[71] In addition to SERS biosensing, these extraordinary electric field enhancements are useful for applications that utilize light-matter interactions, such as harmonic generation, surface enhanced fluorescence, and other applications. Finally, the novel use of EHD flow to control chemical assembly paves the way for increased use of systems invoking EHD flow to fabricate nanoarchitectures composed of assemblies of discrete meta-molecule building blocks from colloid.

## Acknowledgment
The authors acknowledge the National Science Foundation EECS- 1449397 for funding this work. The authors also acknowledge the use of the facilities within the Laser Spectroscopy Facility and the Laboratory for Electron and X-ray Instrumentation (LEXI) center at the University of California, Irvine. The authors would like to thank Shen J. Dillon in the Materials Science and Engineering Department at the University of Illinois at Urbana-Champaign for transmission electron microscopy analysis. The authors would like to thank also Computer Simulation Technology (CST) of America, Inc. for providing CST Microwave Studio that was instrumental in this work.

## MATERIALS AND METHODS

**Materials**

Random copolymer Poly(styrene-co-methyl methacrylate)-α-Hydroxyl-ω-tempo moiety (PS-r-PMMA) ($M_n$ = 7,400, 59.6% PS) and diblock copolymer poly(styrene-*b*-methyl methacrylate) (PS-*b*-PMMA) diblock copolymer PS-*b*-PMMA ($M_n$ = 170-*b*-144 kg mol$^{-1}$) were purchased from Polymer Source, Inc. (Dorval, Canada). Gold nanospheres diameter of 20 nm and 40 nm with lipoic acid functionalization were purchased from Nanocomposix (San Diego, CA). Si(001) wafers with resistivity of 0.004 ohm-cm were purchased from Virginia Semiconductor (Fredericksburg, VA. Hydrofluoric acid (HF) was purchased from Fisher Scientific (Pittsburgh, PA). 2-(*N*-morpholino)ethanesulfonic acid (MES) 0.1M buffer, 1-ethyl-3-[3-dimethylaminopropyl] carbodiimide hydrochloride (EDC), and N-hydroxy sulfosuccinimide (S-NHS) were purchased from Pierce (Rockford, IL). Dimethyl sulfoxide (DMSO), ethylenediamine, benzenethiol, toluene, ethanol, isopropanol (IPA), potassium carbonate, and 52-mesh Pt gauze foil were all purchased from Sigma Aldrich (St. Louis, MO). Nanopure deionized water (DI) (18.2 MΩ cm$^{-1}$) was obtained from a Milli-Q Millipore System.

**Nanoantenna Oligomer Substrate Fabrication**

Lamella PS-b-PMMA block copolymer is spin-coated onto a HF-cleaned, heavily doped Si wafer and annealed as described in previous work[56]. PMMA regions are selectively functionalized with amine end groups by first immersing the entire substrate in DMSO and then in ethylenediamine/DMSO solution (5% v/v), both for 5 minutes without rinsing between steps. The Si substrate coated with functionalized copolymer is then washed with IPA for 1 minute and dried under nitrogen. PS-b-PMMA block copolymer templates on ITO-coated glass are fabricated identically, but using ITO-coated glass that is oxygen plasma etched at 100W for 1 minute instead of the Si wafer.

Lipoic acid functionalized 40 nm [20 nm] Au nanosphere solution is concentrated twofold by adjusting the pH to 8 with potassium carbonate and centrifuging for 25 [30] minutes at 1700 [6500] RCF and redispersed in DI water. 3 mL of concentrated Au nanosphere solution is added to a 10 mL beaker. 35 μL [20 μL] of freshly prepared 20 mM s-NHS in a 0.1 M MES buffer is added to the beaker and swirled. Next, 35 μL [20 μL] of freshly prepared 8 mM EDC in a 0.1 M MES buffer is added to the beaker and swirled. The beaker is placed on a hotplate and brought to 60 $^0$C. A 1 cm x 1 cm functionalized copolymer-coated Si substrate is placed into the solution vertically, held in place as the cathode with alligator clips that do not contact the nanoparticle solution. 5 mm away from the substrate, a 1 cm x 1 cm Pt mesh is placed into the solution vertically, held in place as an anode with alligator clips that do not contact the nanoparticle solution. A DC Regulated Power Supply was used to apply a voltage of 1.2 V for 10 minutes. The substrate, Pt mesh, and beaker are rinsed with IPA for 1 minute and dried under nitrogen. This process is repeated with the same substrate and fresh nanosphere solution as described above, but with 25 μL [12 μL] of EDC and s-NHS solution.

**Characterization**

After nanoantenna oligomers are assembled onto the block copolymer-coated Si substrate, images are collected with a Magellan XHR SEM (FEI), and EOL 2010F STEM using high angle annular dark-field imaging in scanning transmission electron microscopy (STEM) mode.

UV-Vis absorption spectra are taken of nanoantenna oligomer on ITO-coated glass substrates taped (away from the beam path) onto a quartz cuvette. The cuvette is then filled with water and imaged with a Shimadzu UV-1700 absorption spectrometer.

Raman spectroscopy measurements are conducted using a Renishaw InVia micro Raman system with a laser excitation wavelength of 785 nm for 40 nm Au nanosphere samples. For 20 nm Au nanosphere samples, Raman spectroscopy measurements are conducted using a home built Raman microscopy system with a laser excitation wavelength of 633 nm. In both cases, laser excitation wavelength values are chosen based on simulations from previous work. All measurements taken at 152 μW for 10 seconds unless otherwise stated. For measurements of substrates, a 60X water immersion objective with a 1.2 NA, immersed in DI water, is used. Substrates are immersed in 2 mM benzenethiol in ethanol overnight, rinsed with methanol for 1 minute, and dried under nitrogen. Neat measurements are taken by placing a thin layer of BZT between a silicon wafer and class coverslip.

**Simulations**

Full-wave simulations (frequency domain finite elements method solver) are implemented in CST Microwave Studio (CST AG). We simulate absorption and scattering by several nanosphere oligomers: dimer, linear trimer, linear quadrumer, close-packed quadrumer, close-packed hexamer, and close-packed octamer. A schematic of the simulation conditions is presented in Figure 4 a. We consider Au nanospheres with diameter of 40 nm using the Drude model with parameters extracted from Grady et al[72]. We use a 0.9 nm gap between nanospheres, consistent with both observation and the modeled length of an anhydride linker. The nanospheres are shown to be partially embedded in PMMA, with the PMMA layer thickness set to 40 nm, and the center of the nanospheres 8 nm above the layer. Below the PMMA layer is 150 nm of layer of ITO on top of a 2 μm layer of glass. The relative electric permittivities of water, PMMA, gap region, ITO, and glass used in the simulations are 1.77, 2.47, 2.25, 3.3378+i0.011330 and 2.3207+i0.0000011874 respectively. Oligomers are excited with plane wave illumination at normal incidence with electric field polarization along the axis of the linear oligomers, and the absorption cross section of the structure is determined.

**Table of Contents Graphic**

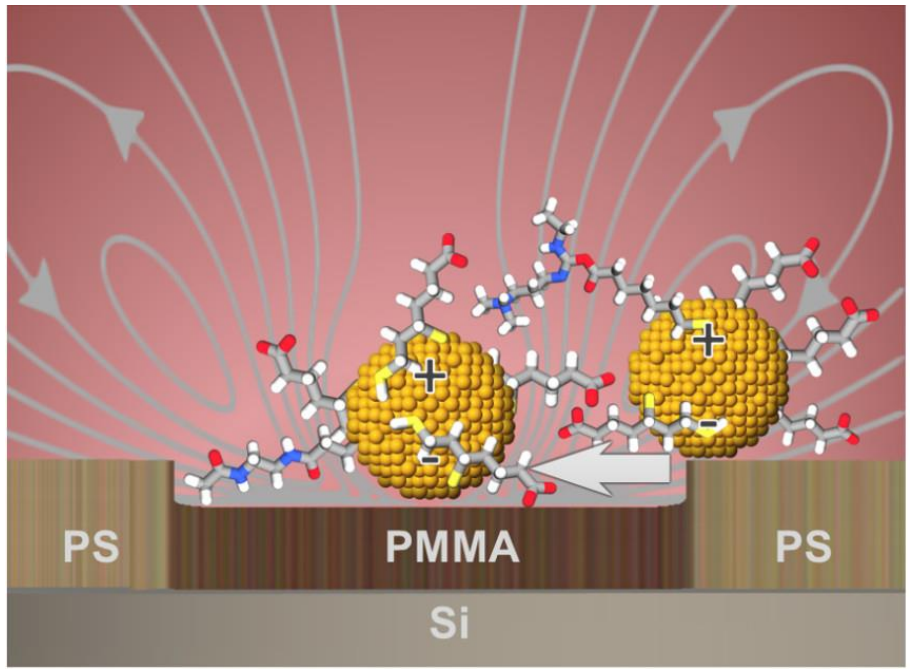
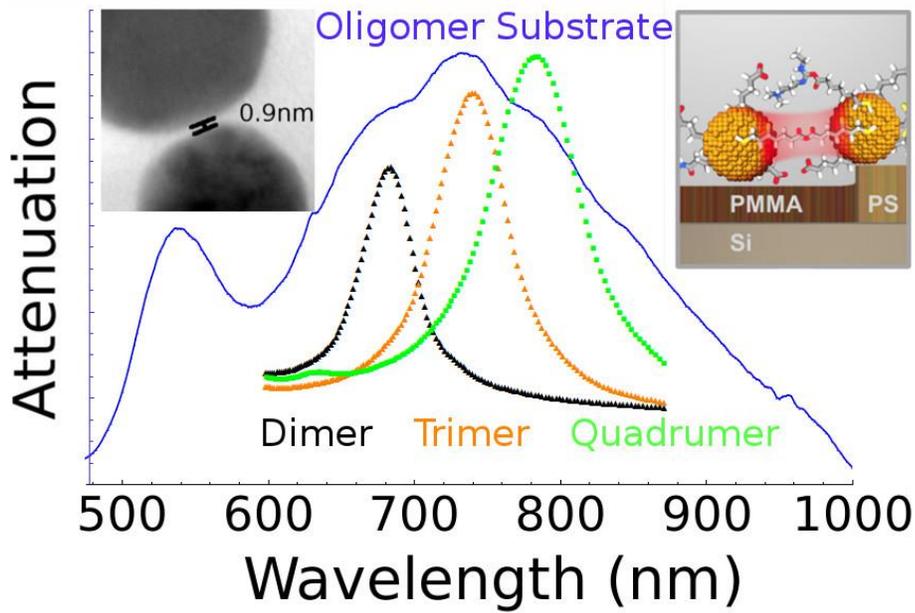

**SUPPLEMENTAL INFORMATION**

To analyze the degree of oligomerization of a given oligomer and the number of nearest neighbors of a given Au nanosphere, Wolfram Mathematica[TM] is implemented for image analysis. SEM images, an example shown in Figure S1 a, are first binarized, and oligomers are distinguished from one another as being separate collections of foreground pixels, or morphological components. Single Au nanospheres are identified through their circularity, defined as the ratio between the equivalent disk perimeter length and the perimeter length of a polygon formed by the centers of each perimeter element, only morphological components over a certain threshold of circularity are determined to be single nanospheres. The remaining morphological components are then divided into component nanospheres using a modified Euclidean distance transform approach, which is designed for implementation in SEM images, where edge effects can make identification of small nanoparticles in close packed structures difficult[69]. For each morphological component, the original image is again binarized using local adaptive binarization, shown in the left side of Figure S1 (b). A Euclidean distance transform (figure distance transform) then reveals the nanoparticle centers as local maxima, shown in the right side of Figure S1 (b). The maxima are then used to determine the center of each nanosphere and a distance threshold is used to determine the number of nearest neighbors of a given nanosphere. From these images the number, density, and nearest neighbor statistics are obtained as shown in Figure S1 (c) and (d), respectively.

Figure S1 (a) SEM image of 40 nm EPD sample (b) left: local adaptive binarized oligomer, right: distance transform of the left image (c) Colorized image of the SEM image with oligomers identified (d) Histogram of the number oligomers identified for each degree of oligomerization.

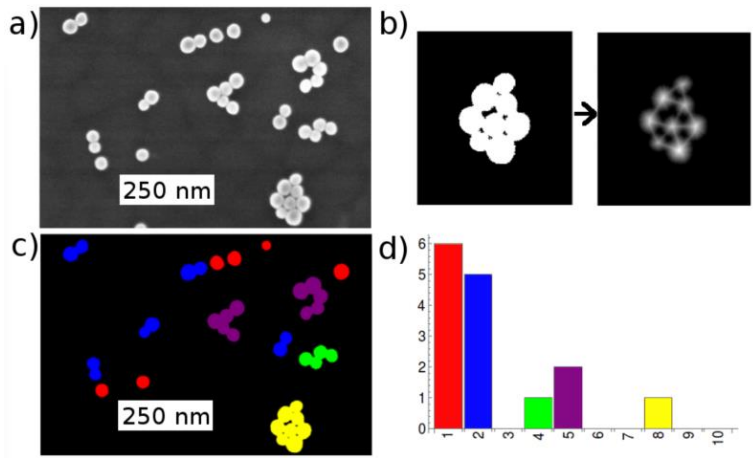

Figure S2 (a) 50 μm x 50 μm SEM image of a 40 nm EPD sample. (b) 2.2 μm x 2.2 μm SEM image of a 40 nm EPD sample.

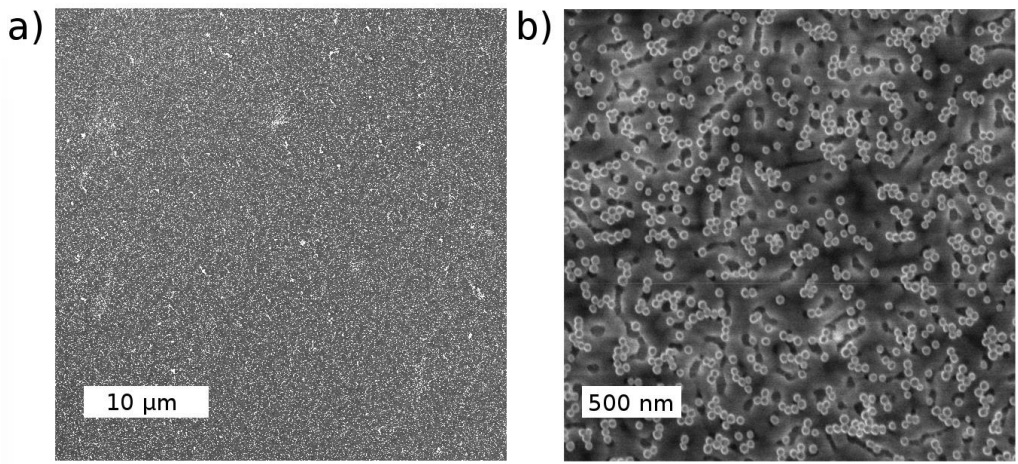

In addition to the SERS data shown in Figure 7, the large-scale uniformity of the nanoantenna oligomers can also be observed from scanning electron microscope (SEM) images. Figure S2 a shows a 50 μm × 50 μm SEM image of a 40 nm EPD sample. The image shows that we do not observe very large-scale aggregates of Au nanospheres, and that oligomer density variations are essentially random and small in scale. Figure S2 b shows a 2.2 μm × 2.2 μm SEM image of a 40 nm EPD sample. From this image one can see the relatively narrow distribution of oligomerization provided by the electrohydrodynamically driven anhydride crosslinking between Au nanospheres. Additionally, one can observe Au nanospheres that are over the PS domains, apparently not linked through a peptide bond to the PMMA domains. The close packing of oligomers is also observed in most of the oligomers in this image.

Figure S3: Field enhancement in the hotspot for oligomers of different geometries

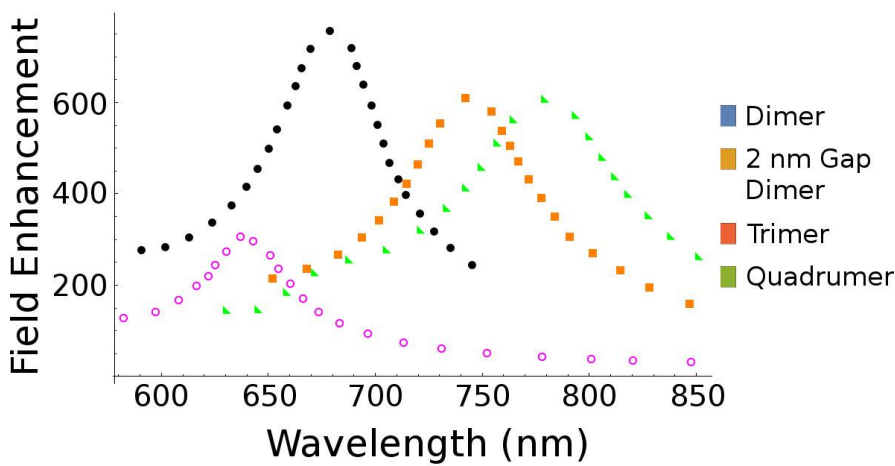

Figure S3 shows the electric field enhancement of oligomers of different geometries. As surface enhanced Raman scattering intensity is based on field enhancement to the fourth power, the predicted enhancement factor (EF) is $3.3 \times 10^{11}$, $1.3 \times 10^{11}$, and $1.3 \times 10^{11}$ for dimer, trimer, and quadrumer respectively. These values correlate well to the average measured EF of $4.5 \times 10^{10}$ for the 40 nm EPD sample. Given the low field enhancement of oligomers with larger gap spacings than 0.9 nm, the length of the anhydride bridge, we see these anhydride bridges play a crucial role in the observed performance of the samples.